\documentclass[%
 aip,
 jmp,%
 amsmath,amssymb,
 reprint,%
]{revtex4-1}

\usepackage{graphicx}
\usepackage{dcolumn}
\usepackage{bm}
\usepackage{setspace}
\usepackage{chngcntr}

\begin{document}

\title{Simulation Study on Local Influence Diagnosis for Poisson Mixed-Effect Linear Model}

\author{N. Zhang}

\affiliation{ 
School of Statistics and Information, Shanghai University of International Business and Economics.
}%

\date{\today}

\begin{abstract}
\textbf{Abstract:} Given that hierarchical count data in many fields are not Normally-distributed and include random effects, this paper extends the Generalized Linear Mixed Models (GLMMs) into Poisson Mixed-Effect Linear Model (PMELM) and do numerical simulation experiments to verify the approach proposed by Rakhmawati et al. (2016) in detecting outliers. This paper produces random data based on epilepsy longitudinal data in Thall and Vail (1990), use six ways to contaminate it and try to use code mentioned in supplementary materials in previous research to detect the man-made outlier. Output shows that this method is effective sometimes but does not always work, this is probably because of the  limitation of coding or some other reasons. Even though the data set and local influence method has been researched and analyzed extensively in previous papers, this paper makes contributions in data visualization. Figures in this paper show the effect of each influencial component, which are clearer than the original output in R and SAS.
\end{abstract}

\maketitle

\section{Introduction}

Count data are frequently encountered in many areas especially in medicine, biomedicine and public health. For example, the epileptic seizures patients have during a period take values 0,1,2\ldots. For a given patient, the counts are often approximately Poisson distributed. Generalized linear models (GLMs; McCullagh \& Nelder, 1989) are frequently used to describe the relationship between counts and one or some variables. It is an extension of linear models in which discrete random variables in exponential family are acceptable, but it requires dependent variable observations of to be independent, which is not true in many situations where some measures are taken from the same individual or subject. This need nurtures the appearance of generalized linear mixed models (GLMMs; Breslow \& Clayton, 1993; Wolfinger and O'Connell, 1993; Engel and Keen, 1994).
GLMM combines the advantages of both LMM and GLM, because dependent variables are not necessary to be Normally-distributed and the model can include random effects. Poisson mixed-effect linear model (PMELM), which this paper focuses on, is a special model of GLMM's.

Some applications of PMELM are as follows:

(1)A research of the application of hierarchical mixture ZIP mixed regression models was done by Chen,Shi and Wang (2016). The data set was obtained from 9 Shanghai middle schools which includes three variables (gender, age and BMI) and the number of push up. Observations for male are more over-dispersed than those of female. In order to fit the data more properly, there assume L latent classes of school level, and for a certain latent class l, $y_{ij}$ (for j-th observation in i-th school) can be described as a ZIP mixed regression model. The aim of hierarchical modeling is to measure the necessity of taking account of unconsidered heterogeneity at school level.

(2)Aiming to find the market structure and forcast brand sales, submarket sales as well as market sales, Terui, Ban and Maki (2009) analyzed hierarchical count data which includes three makers that comprises three brands each, and each maker uses degrees of spiciness to distinguish different categories of products. The data set consists of $y_{it}$(sales for brand i at time t, 110 weeks' sales of these nine brands in total) and a mix variable $X_{it}$which is a mixture of two other variables, price and feature. The article proposes a structural equation and there are three layers marked with superscript s, for s equals 1 to 3, explanatory variable $X_{it}^{s}$ represents market sales, submarket sales and brand sales respectively. In conclusion, the tables in the article reflects that there are three possible market structures: product category, maker, and usage (a criterion defined because of the difference in target use). This three-layer structure model can be extended to describe more complex situations in market.

(3)Samples in studies of microbiome are always with hierarchical structures, while researches about mixed-effect models of microbiome data are really sparse. According to Zhang, Mallick and Tang (2017), it proposes $C_{ij}$ (i-th sample with j-th feature, n samples and m features in total), total sequence read $T_{i}$, environmental or genetic variable $X_{i}$, and sample variable $Z_{i}$, their model can be efficient to solve hierarchical microbiome count data. Based on the assumption that $C_{ij}$ follows a negative binomial distribution, it creates IWLS algorithm to fit NBMM (negative binomial mixed model, one type of GLMM) to find out the relationship between environmental factor and microbiomes and update more parameters. After drawing figures and simulation study, the new proposed method has outperformed and provided an efficient tool to analyze microbiome count data.

In data analysis, there usually exist some observations which are quite different from other observations. If great changes of statistical conclusions take place after perturbing or deleting a certain observation, it is regarded as an influential observation. Obviously, inappropriate diagnostics of influential observations may lead to fallacy in statistical analysis.

Cook (1986) proposed the idea of local influence, the analysis based on likelihood theory and perturbation is widely used nowadays. There are some studies on diagnostics of influential observations in PMELMS. 

The aim of this paper is to study the effectiveness of the existent methods in detecting influential observations. There exist papers focusing on parameter estimate of GLMM, such as Zhu et al.(2003) analyzed data in local influence method, Liang et al.(2006) used EM algorithm to analyze point-deleted model, however, diagnostics still remain less, while Rakmawati et al.(2016) proposed diagnostic method for hierarchical count data with over-dispersed excess zeroes and proved its practicability using in models of Molenberghs et al. (2010). Moreover, present diagnostic plots of SAS do not include decomposition of influential factors, this paper will offer examples for data visualization, in which outliers could be recognized easily. Section 2 are preliminaries, including introduction of PMELM, local influence and diagnosis . Section 3 focuses on simulation study of epileptic data and diagnosis plots. Section 4 is the conclusion part, improvement as well as deficiencies are mentioned.

\section{Preliminaries}

\subsection{Poisson Mixed Effect Linear Models (PMELM)}

Linear mixed model (LMM) is given by
\begin{equation}
\boldsymbol{y}=\boldsymbol{X\beta}+\boldsymbol{Zu}+\boldsymbol{\alpha}
\end{equation}
where $\bm{y}$ is a $n\times1$ vector of dependent variables, matrix $\bm{X}$ is a  $n\times p$ matrix of covariants, $\bm{\beta}$ is a $p\times1$ vector of fixed effects, and $\bm{Z}$ is a $n\times q$ are known incidence matrix, respectively. $\bm{u}$ is a $q\times1$ vector of random effects, $\bm{\alpha}$ is a  $n\times1$ vector of random errors. Moreover, we assume that $E(\bm{u})=\textbf{0}, E(\bm{\varepsilon})=\textbf{0}$ and 
\begin{equation}
Var\left( 
\begin{matrix}
\bm{u}\\
\bm{\alpha}
\end{matrix}\right) 
=\left( 
\begin{matrix}
\bm{G}& \bm{0}\\
\bm{0}& \bm{R}
\end{matrix}\right), 
\end{equation}
where $\bm{G}$ is positive semidefinite and $\bm{R}$ is positive definite. Parameters $\bm{\beta}$, $\bm{G}$ and $\bm{R}$ are typically unknown.

The PMELMs can be defined as follows. Let $y_{ij}$ be the $jth$ measurement of the $ith$ subject, we assume
\begin{equation}
	y_{ij}|(\mu_i,\alpha_j)\sim Poi(\mu_{ij})
\end{equation}
\begin{equation}
log(\mu_{ij})=\eta_{ij},i=1,\cdots,m_1,j=1,\cdots,m_2,
\end{equation}
\begin{equation}
	\eta_{ij}=x_{ij}^T\beta+u_i+\alpha_j
\end{equation}
$\newline u=(u_1,\cdots,u_{m_1})^T\sim N\{0_{{m_1}\times1},\sigma_1^2I_{m_1}\},\newline  \eta=X\beta+Zu+\alpha=(\eta_{11},\cdots,\eta_{{m_1},{m_2}})^T$ where elements are ordered exicographically\newline $X=(x_1^T,\cdots,x_{1,{m_2}}^T,\cdots,x_{{m_1},1}^T,\cdots,x_{{m_1},{m_2}}^T),$ \newline $Z=(I_{m_1}\otimes\bm{1}_{m_2}\quad \bm{1}_{m_1}\otimes I_{m_2})$

In PMELM, $\beta$ is fixed effects, $u_i$ is random effect with Normal distribution assumption, $X$ and $Z$ are design matrix. $\alpha_j$ is the error term.

\subsection{Local Influence and Perturbations}

Local influence was proposed by Cook (1986). Cook and Weisberg (1982) as well as Chatterjee and Hadi (1988) used the idea of local influence, which means that measuring the fluctuation of a certain statistical parameter through case deletion. A convenient form of $C_i$ could be given by
\begin{equation}
C_i=-2{(\bm{\hat{\theta}}-\bm{\hat{\theta}}^{1}_{(i)})}^{'}\ddot{L}_{(i)}\ddot{L}^{-1}\ddot{L}_{(i)}(\bm{\hat{\theta}}-\bm{\hat{\theta}}^{1}_{(i)})
\end{equation},
where $i$ means that the corresponding number is because the deletion of the $i$th subject and $\bm{\hat{\theta}}^{1}$ is the one-step approximation of $\bm{\hat{\theta}}$, which is obtained from a single Newton-Raphson step in the maximization process of $l_{(i)}(\bm{\theta})$, starting from $\bm{\hat{\theta}}$.
Cook(1986) and Beckman et al.(1987) was further developed by Lesaffre and Verbeke (1997b and 1998) in LMM, in which the estimates was measured by the impact of case-weight perturbation. Let the log-likelihood of GLMM or its combined extension  is in the form 
\begin{equation}
l(\theta)=\sum_{i=1}^Nl_i(\theta)
\end{equation}
where $l_i(\theta)$ denotes the contribution of i-th subject to the log-likelihood. Let
\begin{equation}
l(\theta|\omega)=\sum_{i=1}^N\omega_il_i(\theta)
\end{equation}
represent the perturbed version of $l(\theta)$, which relies on an N-dimensional vector $\omega$ of weights and assumed to belong to an open subset $\Omega$ of $R^N$. The original log-likelihood follows for $\omega=\omega_0=(1,1,\ldots,1)^{'}.$
Rakmawati et al.(2016) applies it in GLMM, the local influence for the Probit-normal Model could be interpreted by fixed effect and squared of variability of random effect.
For zero-inflated count models, there are several studies like Xie et al. (2008) and Chen et al. (2013) who use local influence in zero-inflated generalized Poisson mixed-effect models. Garay et al.(2011) took local influence under various perturbation schemes for zero-inflated binomial models.

\subsection{Influence Diagnostics of PMELM}

According to Lesaffre and Verbeke (1998), local influence $C_i$ could be interpreted by five sepertate components, $\lVert \mathcal{X}_i{\mathcal{X}_i}^{'} \rVert$, $\lVert \mathcal{R}_i \rVert$, $\lVert \mathcal{Z}_i{\mathcal{Z}_i}^{'} \rVert$, $\lVert I- \mathcal{R}_i{\mathcal{R}_i}^{'} \rVert$, and $\lVert V_i^{-1} \rVert$. $r_i=y_i-X_i\hat{\beta}$, $\lVert r_i{r_i}^{'} \rVert$ is an estimate of var($y_i$). Here in Poisson-normal model,  
\begin{equation}
C_i=2\lVert\ddot{L}^{-1}\rVert\lVert\Delta_i\rVert^2cos(\varphi_i)
\end{equation}
where
\begin{equation}
\lVert\Delta_i\rVert^2=(\sum_{j=1}^{n_i}r_{ij}x_{ij})(\sum_{j=1}^{n_i}r_{ij}x_{ij})^{'}
\end{equation}
$C_i$ could be seperated into two parts, $C_{1i}$ and $C_{2i}$. 
\begin{equation}
C_{1i}=2\lVert\ddot{L}^{-1}\rVert\lVert x_ix_i^{'}\rVert\lVert r_i\rVert^2cos(\alpha_i)cos(\varphi_i)
\end{equation}
\begin{equation}
\begin{aligned}
C_{2i}&=\frac{1}{2}\lVert\ddot{L}^{-1}\rVert cos(\varphi_i) 
\\ &\times[tr\{{(D^{-1})_{kl}^2}\}-tr\{2{(D^{-1})_{kl}(D^{-1}D^{-1})_{kl}Var(\bm{b_i})}\}
\\
&+tr\{(D^{-1}D^{-1})_{kl}^2Var(\bm{b_i})^2\}]
\end{aligned}
\end{equation}
$\lVert x_ix_i^{'}\rVert$ is the 'length of fixed effect', $\lVert r_i\rVert^2$ is the squared 'length of the residual', $\alpha_i$ is the angle between vector $x_ix_i^{'}$ and vector $r_ir_i^{'}$, $\varphi_i$ is the angle between vector $-\ddot{L}^{-1}$ and vector $\Delta_i\Delta_i^{'}$, $Var(\bm{b_i})^2$ is the 'squared of random effect variability'.
In this paper, we use $C_i\_b$\ and\ $C_i\_d$\ to represent $C_{1i}$ and $C_{2i}$\  respectively, $C_i\_b$\ represents fixed-effects and $C_i\_d$\ represents random-effects.

\section{Simulation Experiments}

\subsection{Introduction and Application to Epileptic Dataset}

This section use data set \emph{epil} in R to do simulation analysis. This data set was analyzed in Molenberghs et al. (2007) which included Gamma random effects in the Poisson-Normal Model, and Kassahun et al. (2014) extended it into ZIPGN and HPGN model. In the research of epidemiology, to study the efficacy of progabide, 59 patients are randomly assigned into treatment group or control group, and for each group, the number of epileptic seizures are recorded in a period of 8 weeks which is divided into 4 successive two-week periods. $y_i$ is the counts of disease in 2-week period, $trt$ stands for treatment, where $trt=0$ means ``placebo" and $trt=1$ means ``progabide". ``base" is the frequency in the baseline 8-week period, ``lbase" is the centralized log-counts for baseline period and ``lage" is the centralized log-age for each subject.

\subsection{Design of Contamination}

 Based on the previous PMELM, find maximum likelihood estimate of $X,\beta\ and\ Z$  by using glmmPQL in R, and then random data $Y_i,\ (i=1,2,3)$ with different $\sigma_{1i}$ was generated. Each \bm{$Y_i$} is a matrix of $59\times 5$ degree, \bm{$Y_{i1}$} is the simulation for baseline, \bm{$Y_{i2}$} to \bm{$Y_{i5}$} are simulations for counts of certain 2-week period.  $\sigma_{1i}=0.25,\ 0.5\ and\ 1$ respectively. Histograms of sum of seizures in original data and three random data sets are shown in Figure 1. Now use 6 methods to contaminate the random data to change $y_{1}$ into an outlier, as is shown in Table 1. Contaminated data are shown in Figure 2 to 4, red line represents using different ways to change the same patient's observation, the other five black lines represents observations of another 5 patients, which are selected randomly. However,  Because the code doesn't include influence of baseline, so the above 6 methods could shrink to 4, which means that method 5 and 6 has the same output with method 1 and 4, so only the first 4 methods are discussed in the following part.

\begin{table}[!htp]
	\centering
	\caption{Data Contamination Methods}
	\begin{tabular}{c|p{5cm}}
		\hline
		id& method\\
		\hline
		(1) & Add 30 to $y_{12}$\\
		(2) & Add 30 to $y_{12}$, $y_{13}$, $y_{14}$ and $y_{15}$\\
		(3) & Add 100 to $y_{12}$\\
		(4) & Add 100 to $y_{12}$, $y_{13}$, $y_{14}$ and $y_{15}$\\
		(5) & Change $y_{11}$ into 50\\
		(6) & Add 100 to $y_{12}$, $y_{13}$, $y_{14}$ and $y_{15}$, and change $y_{11}$ into 50\\
		\hline
	\end{tabular}
\end{table}

\begin{figure}[!ht]
	\centering
	\includegraphics[width=3.3in]{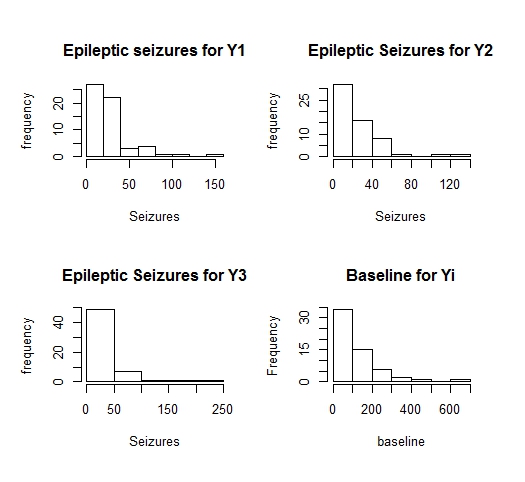}
	\caption{Epileptic Seizures for \bm{$Y_{i}$}}
\end{figure}

\begin{figure}[!ht]
	\centering
	\includegraphics[width=3.3in]{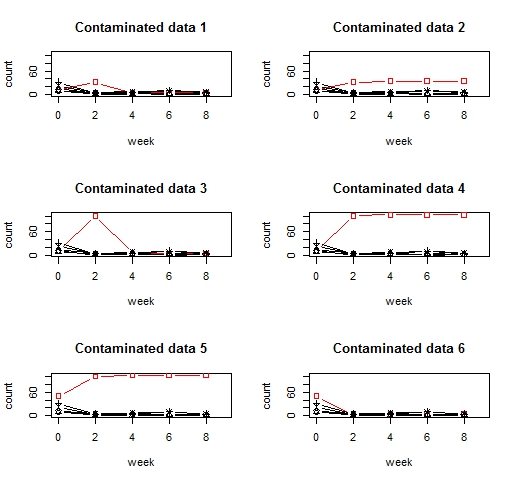}
	\caption{Contaminated Data with $\sigma_1=0.25$}
\end{figure}

\begin{figure}[!ht]
	\centering
	\includegraphics[width=3.3in]{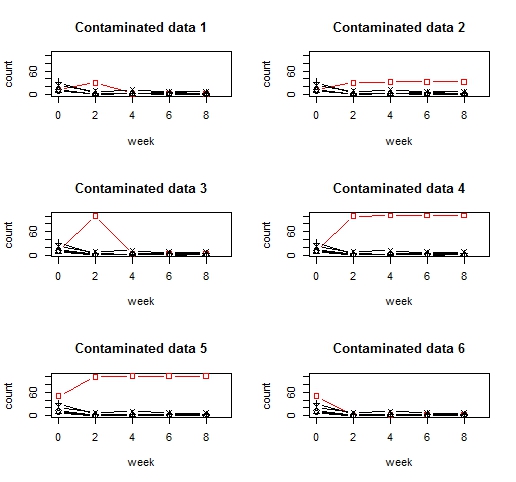}
	\caption{Contaminated Data with $\sigma_1=0.5$}
\end{figure}

\begin{figure}[!ht]
	\centering
	\includegraphics[width=3.3in]{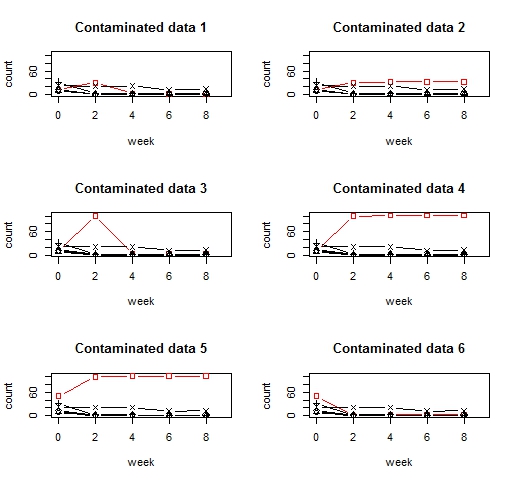}
	\caption{Contaminated Data with $\sigma_1=1$}
\end{figure}

\subsection{Diagnostic Plots}

 This subsection uses methods put forward in Rakhmawati et al. (2016) to try to diagnose man-made outliers. Fixed and random effects($C_i$) could be decomposed into two parts, $C_i\_b$\ and\ $C_i\_d$. Scatter plots of $C_i\_d$ and $C_i\_b$, needle plots of $C_i$, $C_i\_b$,  $C_i\_d$, and $rr_i$ are given as follows. For clear visualization in needle plots,  only 20 dots,10 with $trt$=0 and 10 with $trt$=1 are randomly selected to show the feasibility of diagnosis.

\begin{figure}[!ht]
	\centering
	\includegraphics[width=3.3in]{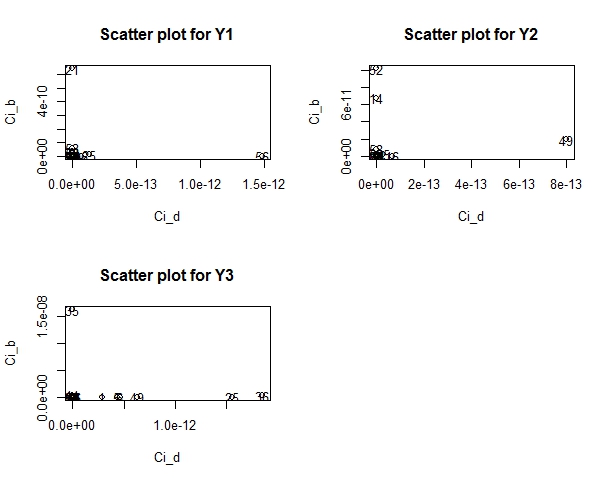}
	\caption{Scatter Plots for \bm{$Y_{i}$} }
\end{figure}

In scatter plots, in \bm{$Y_{1}$}, $y_{21}$ only have fixed effects, $y_{56}$ have random effects; in \bm{$Y_{2}$},  $y_{14}$ and $y_{52}$ only have fixed effects; in \bm{$Y_{3}$}, $y_{35}$ only have fixed effects, $y_{25}$ and $y_{36}$ only have random effects.

\begin{figure}[!ht]
	\centering
	\includegraphics[width=3.3in]{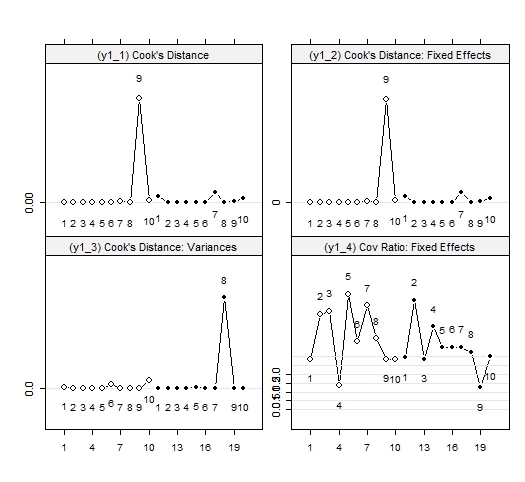}
	\caption{Diagnostic Plots for  \bm{$Y_{1}$}}
\end{figure}

\begin{figure}[!ht]
	\centering
	\includegraphics[width=3.3in]{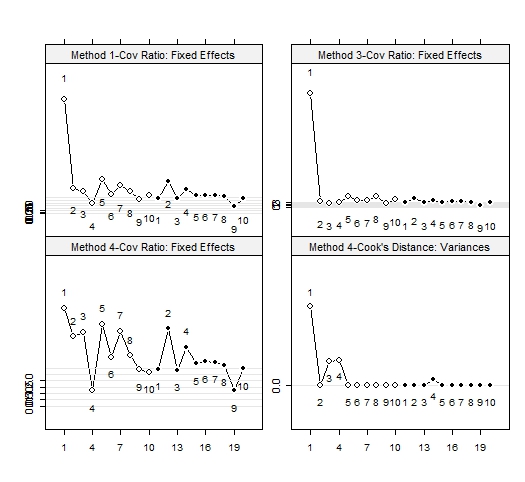}
	\caption{Diagnostic Plots for Contaminated \bm{$Y_{1}$}}
\end{figure}

Figure 6-11 are output of simulation diagnosis. Figure 6, 8, 10 are outlier diagnosis for 3 random data sets without contamination. Figure 7, 9, 11 are summary of methods and parameters which are efficient to diagnose man-made  outliers in 3 contaminated data sets respectively. Complete output could be seen in Appendix. 

For \bm{$Y_{1}$}, the 9th observation in $trt=0$ and the 8th observation in $trt=1$ are outliers. After contamination, the first observation turns to be an outlier. In Figure 7, $rr_i$ in Method 1, 3, 4 and $C_i\_d$ in Method 4 clearly figure out the first observation as an outlier, the other 12 figures fails to correctly point out the outlier.
For \bm{$Y_{2}$}, the 5th observation in $trt=0$ and the 5th as well as the 6th observation in $trt=1$ are outliers. After contamination, as can be seen in  Figure 9, $rr_i$ in Method 1, 3 and $C_i\_d$ in Method 2,4 clearly figure out the first observation as an outlier, the other 12 figures fails to correctly point out the outlier.
For \bm{$Y_{3}$}, the 2nd observation in $trt=1$ is probably an outlier. In Figure 11, $rr_i$ in Method 1, 3 and $C_i\_d$ in Method 1,2 clearly figure out the first observation as an outlier, the other 12 figures fails to correctly point out the outlier.

\begin{figure}[!ht]
	\centering
	\includegraphics[width=3.3in]{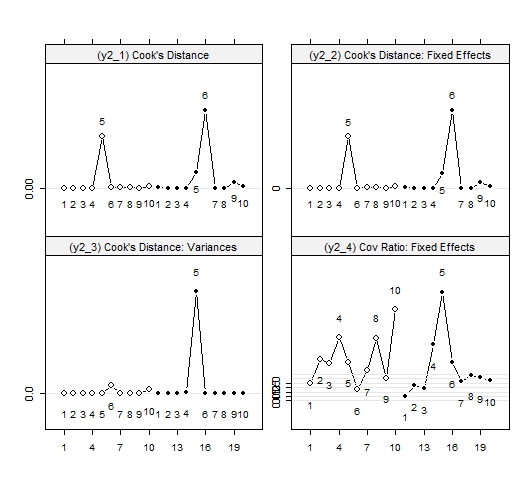}
	\caption{Diagnostic Plots for  \bm{$Y_{2}$}}
\end{figure}

\begin{figure}[!ht]
	\centering
	\includegraphics[width=3.3in]{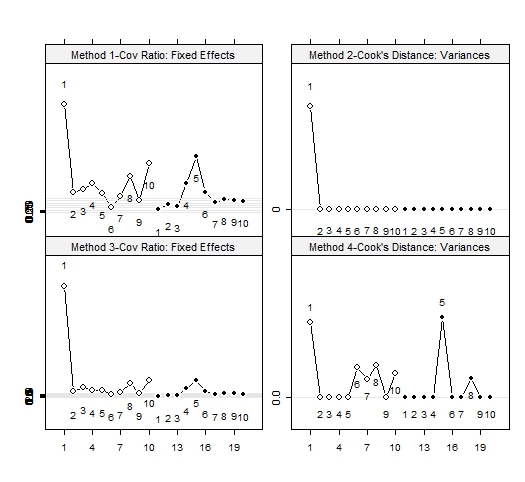}
	\caption{Diagnostic Plots for Contaminated \bm{$Y_{2}$}}
\end{figure}

\begin{figure}[!ht]
	\centering
	\includegraphics[width=3.3in]{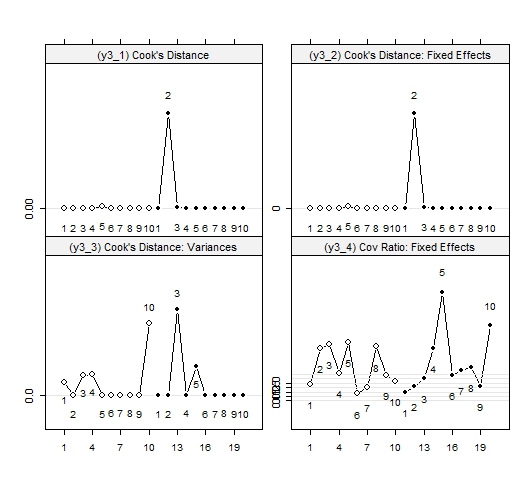}
	\caption{Diagnostic Plots for  \bm{$Y_{3}$}}
\end{figure}

\begin{figure}[!ht]
	\centering
	\includegraphics[width=3.3in]{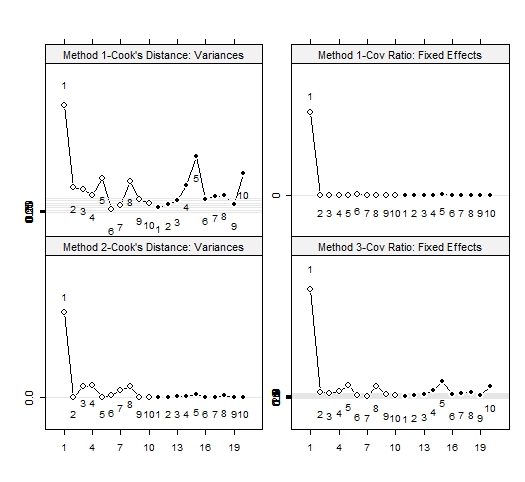}
	\caption{Diagnostic Plots for Contaminated \bm{$Y_{3}$}}
\end{figure}

In all \bm{$Y_{i}$}, one third diagnosis plots could detect the outlier correctly, the reason why this phenomenon occurs may because:

(1)Randomly produced \bm{$Y_{i}$} already contains outliers, they become an interference to the detection of man-made outlier. In all \bm{$Y_{i}$}, $y_{40}$ contains 4 zero observations while  $y_{49}$ has 4 observations high above average epileptic seizures, these may counteract the influence of $y_{1}$ in simulation process.

(2)There maybe exist some faults in the SAS code from Rakhmawati et al. (2016) because Cook distance of man-made outlier become smaller under some circumstances. 

Moreover, from the needle plots, $Ci\_d$ and $rr_i$ are two statistical parameters that detect outliers more precisely in the simulation process.

\section{Conclusion}

It could be seen that in the case of epileptic data, the influential observations for PMELM could be detected by the local influence method in some circumstances, but it does not work all the time. There are some possible reasons which lead to this phenomenon, such as the original outliers' existence and the wrongly coding. In addition, the diagnostic statistical parameters like $Ci\_b$, $Ci\_d$ and $rr_i$ could be used to evaluate the influence, they are obvious and effective under different simulation methods. Moreover, this paper puts forward many useful diagnostic plots that SAS and R's original output couldn't show, data visualization is helpful in figuring out certain effects and the outliers, which is innovative.

For the local influence method, baseline effects could be considered in the future research, which will possibly lead to better detection of outliers.

\section{Reference}

\noindent Beckman, R. J., Nachtsheim, C. J. and Cook, R. D. (1987). Diagnostics for mixed-model analysis of variance. $Technometrics$ 29:413每426.

\noindent Breslow, N. E. and Clayton, D. G. (1993). Approximate inference in generalized linear mixed models. $Journal\ of\ the\ American\ Statistical\ Association$, 88:9每25.

\noindent Chen, X.D., Fu, Y.Z. and Wang, X.R. (2013). Local influence measure of zero-inflated generalized Poisson mixture regression models. $Statistics in Medicine$32:1294每1312.

\noindent Chen, X.D., Shi, H.X. and Wang, X. R. (2016). Hierarchical Mixture Models for Zero-inflated Correlated Count Data. $Acta\ Mathematicae\ Applicatae\ Sinica,\ English\ Series$, 32:373-384

\noindent Cook, R. D. (1986). Assessment of local influence. $Journal\ of\ the\ Royal\ Statistical\ Society$, Series B 48:133-169.

\noindent Cook, R. D. and Weisberg, S. (1982). Residuals and Influence in Regression. $Chapman\ \& Hall/CRC$, London, UK.

\noindent Engel, B. and Keen, A. (1994). A simple approach for the analysis of generalized linear mixed models. $Statistica Neerlandica$ 48:1每22.

\noindent Garay, A. M, Hashimoto, E. M, Ortega, E. M. and Lachos, V. H. (2011). On estimation and influence diagnostics for zero-inflated negative binomial regression models. $Computational Statistics and Data Analysis$ 55:1304每1318.

\noindent Kassahun, W., Neyens, T., Molenberghs, G., Faes, C. and Verbeke, G. (2014). A zero-inflated overdispersed and hierarchical Poisson model. $Statistical Modelling$ 14:439每456.

\noindent Laird, N. M. and Ware, J. H. (1982). Random-effects models for longitudinal data. $Biometrics$ 38:963-974.

\noindent Lesaffre, E. and Verbeke, G. (1998). Local influence in linear mixed models. $Biometrics$ 54:570每582.

\noindent McCullagh, P. and Nelder, J. A. (1989). $Generalized\  Linear\  Models$. Chapman \& Hall/CRC, London, UK.

\noindent Molenberghs, G., Verbeke, G. and Demetrio, C. (2007). An extended random-effects approach to modeling repeated, overdispersed count data. $Lifetime Data Analysis$. 13:513每531.

Molenberghs, G., Verbeke, G., Demetrio, C. G. B. and Vieira, A. (2010). A family of generalized linear models for repeated measures with normal and conjugate random effects. $Statistical Science$. 25:325每347.

\noindent Rakhmawati, T.W., Molenberghs, G., Verbeke, G. and Faes, C. (2016). Local influence diagnostics for hierarchical count data models with overdispersion and excess zeros. $Biometrical\ Journal$. 00:1-19.

\noindent Terui, N., Ban, M. and Maki, T. (2010). Finding market structure by sales count dynamics〞Multivariate structural time series models with hierarchical structure for count data. $Ann\ Inst\ Stat\ Math$. 62:91-107.

\noindent Thall, P. F. and Vail, S. C. (1990). Some covariance models for longitudinal count data with overdispersion. $Biometrics$. 46:657-671.

\noindent Wolfinger, R. and O＊Connell, M. (1993). Generalized linear mixed models: a pseudo-likelihood approach. $Journal of Statistical Computation and Simulation$ 48:233每243.

\noindent Xie, F., Wei, B. and Lin, J. (2008). Assessing influence for pharmaceutical data in zero-inflated generalized Poisson mixed models. $Statistics in Medicine$ 27:3656每3673.

\noindent Zhang, X.Y., Mallick H., Tang, Z. X., Zhang, L., Cui, X. Q., Benson, A. K. and Yi, N. J. (2017). Negative binomial mixed models for analyzing microbiome count data. $BMC\ Bioinformatics$, 18:4.

\noindent Zhu, H. T. and Lee, S. Y. (2003). Local Influence for generalized linear mixed models. $Canadian\ Journal\ of\ Statistics$, 31:293-309.

~\\ㄩ

\noindent\textbf{Acknowledgement: Financial support from ``2017 Shanghai University Students Innovation and Entrepreneurship Training Program Model School", Shanghai Education Commission.}

\setcounter{figure}{0}
\begin{figure*}
	\begin{minipage}{0.48\linewidth}
		\centerline{\includegraphics[width=5.0cm]{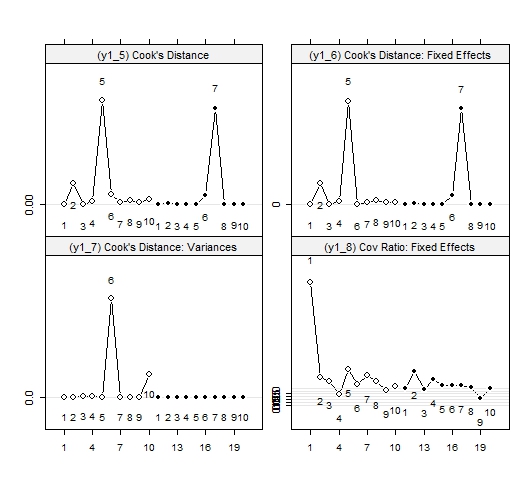}}
		\centerline{Method 1}
	\end{minipage}
	\hfill
	\begin{minipage}{.48\linewidth}
		\centerline{\includegraphics[width=5.0cm]{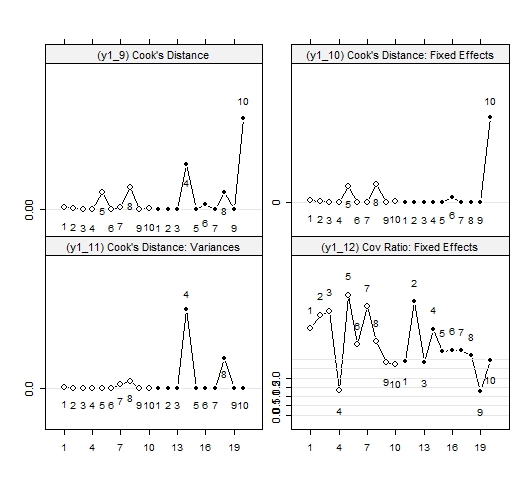}}
		\centerline{Method 2}
	\end{minipage}
	\vfill
	\begin{minipage}{0.48\linewidth}
		\centerline{\includegraphics[width=5.0cm]{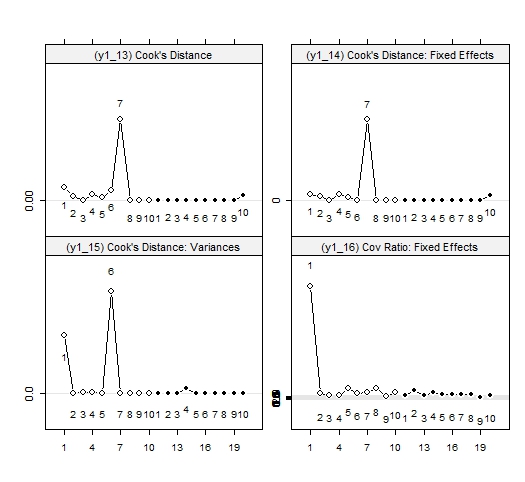}}
		\centerline{Method 3}
	\end{minipage}
	\hfill
	\begin{minipage}{0.48\linewidth}
		\centerline{\includegraphics[width=5.0cm]{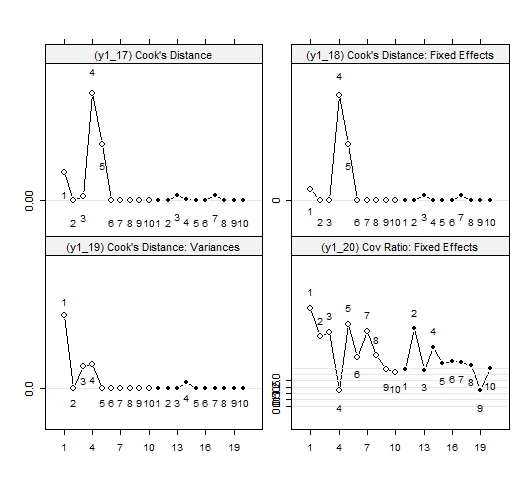}}
		\centerline{Method 4}
	\end{minipage}
	\caption{Diagnostic Plots for  Contaminated \bm{$Y_{1}$}}
	\label{fig:res}
\end{figure*}

\begin{figure*}
	\begin{minipage}{0.48\linewidth}
		\centerline{\includegraphics[width=5.0cm]{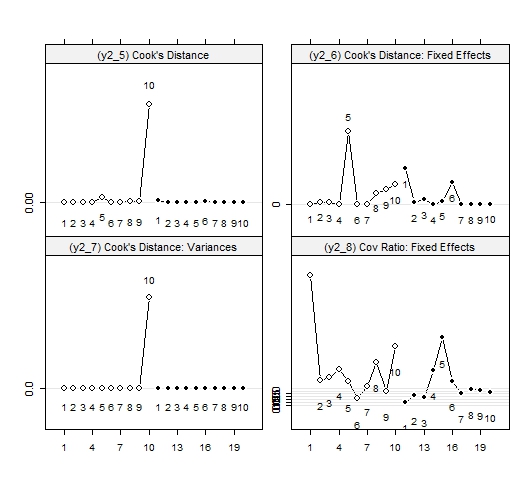}}
		\centerline{Method 1}
	\end{minipage}
	\hfill
	\begin{minipage}{.48\linewidth}
		\centerline{\includegraphics[width=5.0cm]{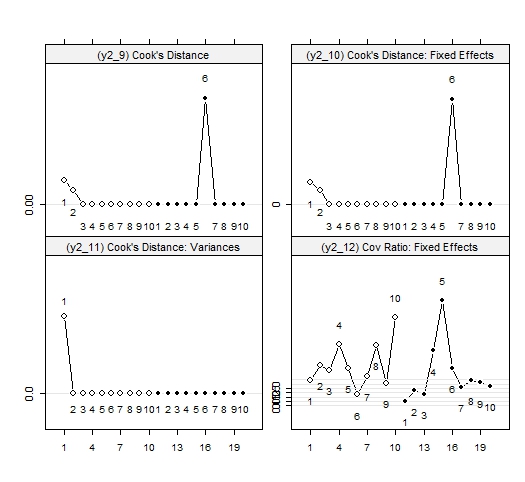}}
		\centerline{Method 2}
	\end{minipage}
	\vfill
	\begin{minipage}{0.48\linewidth}
		\centerline{\includegraphics[width=5.0cm]{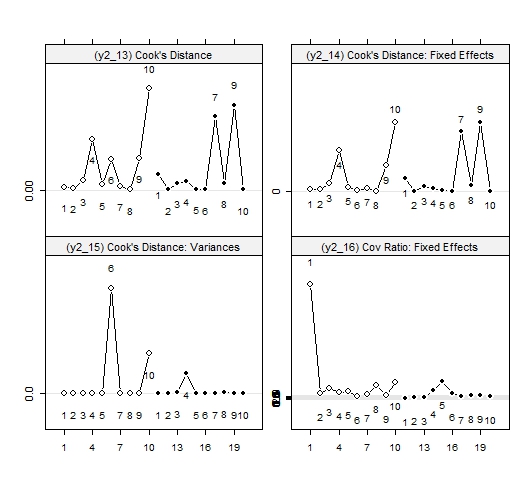}}
		\centerline{Method 3}
	\end{minipage}
	\hfill
	\begin{minipage}{0.48\linewidth}
		\centerline{\includegraphics[width=5.0cm]{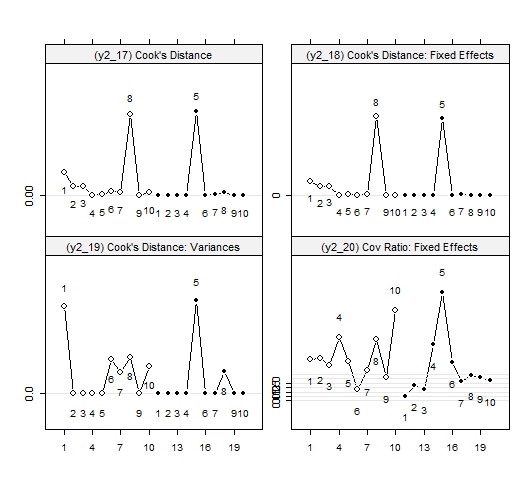}}
		\centerline{Method 4}
	\end{minipage}
	\caption{Diagnostic Plots for  Contaminated \bm{$Y_{2}$}}
	\label{fig:res}
\end{figure*}

\begin{figure*}
	\begin{minipage}{0.48\linewidth}
		\centerline{\includegraphics[width=5.0cm]{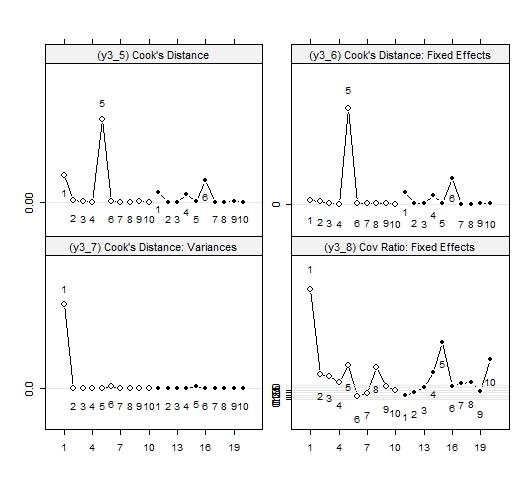}}
		\centerline{Method 1}
	\end{minipage}
	\hfill
	\begin{minipage}{.48\linewidth}
		\centerline{\includegraphics[width=5.0cm]{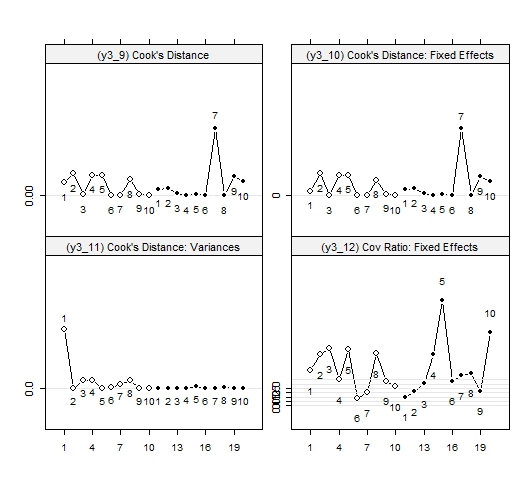}}
		\centerline{Method 2}
	\end{minipage}
	\vfill
	\begin{minipage}{0.48\linewidth}
		\centerline{\includegraphics[width=5.0cm]{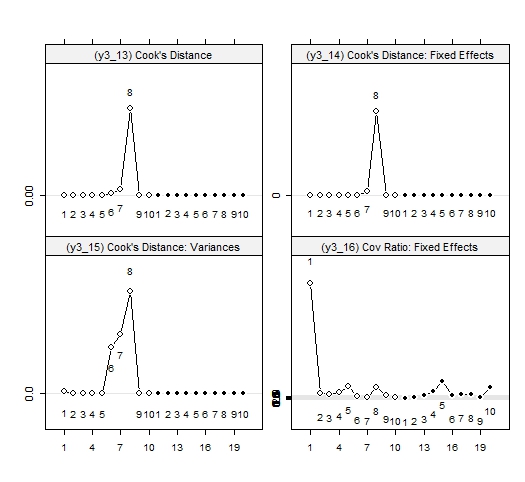}}
		\centerline{Method 3}
	\end{minipage}
	\hfill
	\begin{minipage}{0.48\linewidth}
		\centerline{\includegraphics[width=5.0cm]{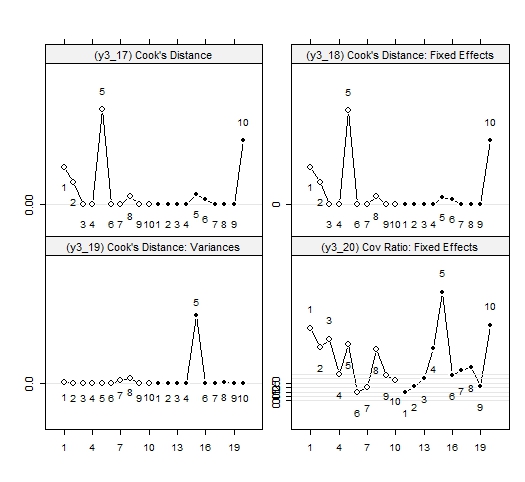}}
		\centerline{Method 4}
	\end{minipage}
	\caption{Diagnostic Plots for  Contaminated \bm{$Y_{3}$}}
	\label{fig:res}
\end{figure*}

\end{document}